\documentstyle[12pt,a4,here,epic,eepic]{article}

     \def\11{1\mbox{\hspace{-0.9ex}}1}

\makeatletter
\@addtoreset{equation}{subsection}

\makeatother

\makeatletter
\long\def\@makecaption#1#2{%
   \vskip 10\p@
   \setbox\@tempboxa\hbox{#1\ \ #2}%
   \ifdim \wd\@tempboxa >\hsize
   #1\ \ #2\par        %
      \else
   \hbox to\hsize{\hfil\box\@tempboxa\hfil}%
   \fi}
\def\fnum@figure{Fig. \thefigure}
\makeatother

\begin{document}

\begin{flushright}
\begin{tabular}{l}
HUPD-9517 \hspace{1em}\\
July 1995
\end{tabular}
\end{flushright}

\vspace{1.5cm}

\begin{center}
{\Large \bf
 Dynamical Symmetry Breaking\\[4mm]
in Einstein Universe}\\[2cm]
\normalsize
K.~Ishikawa
\footnote{e-mail : ishikawa@theo.phys.sci.hiroshima-u.ac.jp},
T.~Inagaki
\footnote{e-mail : inagaki@theo.phys.sci.hiroshima-u.ac.jp}
and T.~Muta
\footnote{e-mail : muta@fusion.sci.hiroshima-u.ac.jp}
\\[1cm]
{\it
Department of Physics, Hiroshima University, \\
Higashi-Hiroshima, Hiroshima 739, Japan \\[3cm]
}
\end{center}

\begin{abstract}
\vglue 0.7cm
We investigate four-fermion interactions
with $N$-component fer\-mi\-on in Einstein universe
for arbitrary space-time dimensions ($2 \leq D<4$).
It is found that the effective potential for composite
operator $\overline{\psi}\psi$ is calculable in the
leading order of the $1/N$ expansion.
The resulting effective potential is analyzed by varying
 the curvature of the space-time and is found to exhibit the symmetry
 restoration through the second-order phase transition.
The critical curvature at which
 the dynamical fermion mass disappears is
analytically calculated.
\end{abstract}

\newpage


\renewcommand{\arraystretch}{2}
\renewcommand{\thesubsection}{\arabic{subsection}}
\renewcommand{\thesubsubsection}
   {\arabic{subsection}.\arabic{subsubsection}}
\baselineskip=24pt

It is the standard scenario of the cosmology to assume that in the
stage of the very early universe the grand unified theory (GUT)
 branched off into the quantum chromodynamics and
electroweak theory through the symmetry breaking caused by the
Higgs mechanism. At this stage the quantum effect of the
gravity is considered to be unimportant although the curvature
effect due to the strong gravity still remains to be  of major
importance.
It is thus of interest to deal with quantum field theory
in curved space-time in the era of the grand unified theory.
If we take the view of the dynamical symmetry breaking
 \cite{NJL} such as the technicolor model \cite{TC},
 the Higgs particle playing an important
role in the symmetry breaking in the GUT era is thought of as
a composite system of the fundamental fermion and anti-fermion.
 Hence it is very interesting to discuss quantum field theory
 with the composite Higgs field in curved space-time \cite{Curv,ESY,TTS}.
 In the present paper we deal with the four-fermion interaction
theory  as one of the prototypes of the composite Higgs theory.
In order to get an insight into the dynamical symmetry breaking
 through the composite Higgs mechanism we study the phase structure
 of the four-fermion theory in curved space-time.

The four-fermion interaction in curved space-time is characterized
 by the action
\begin{equation}
  S=\int d^{D}x \,\sqrt{-g}
  \left[-\sum^{N}_{k=1}
  \overline{\psi}_{k}\gamma^{\mu}\nabla_{\mu}\psi_{k}
  + \frac{\lambda_{0}}{2N}
  (\sum^{N}_{k=1} \overline{\psi}_{k} \psi_{k})^{2}\right] ,
 \label{eqn:action}
\end{equation}
where index $k$ represents the flavor of the fermion,
$N$ is the number of the fermion species,
$g$ the determinant of the metric $g^{\mu\nu}$,
$\gamma^{\mu}$ the Dirac matrix in curved space-time,
$\nabla_{\mu}$ the covariant derivative for fermion field $\psi$,
 and $D$ the dimension of the space-time.
We work in the space-time of dimension $D$ for $2\leq D<4$.
In the following for simplicity we neglect the summation symbol
on indices of fermion field $\psi$.
 The action of Eq.(\ref{eqn:action}) represents a discrete chiral
 symmetry in even dimensions.
It is well-known that in the flat space-time (Minkowski space-time)
 the symmetry is broken spontaneously and the fermion acquires a
 dynamical mass \cite{GN,DGN}.

By introducing auxiliary field $\sigma$ we derive action $S'$
equivalent to action $S$ in Eq.(\ref{eqn:action}):
\begin{equation}
  S'=\int d^{D}x \,\sqrt{-g}
  \left[-\overline{\psi}\gamma^{\mu}(\nabla_{\mu}+\sigma)\psi
  - \frac{N}{2\lambda_{0}}\sigma^{2} \right].
 \label{eqn:action2}
\end{equation}
We perform the path integration over $\psi$ and $\overline{\psi}$
 in the expression for the generating functional and then
 find that in the leading order of the $1/N$ expansion the
 effective action of this model is given by
\begin{equation}
  \Gamma [\sigma] = \int d^{D}\! x \, \sqrt{-g}
  (-\frac{\sigma^{2}}{2\lambda_{0}})
  - i \mbox{Tr}\mbox{Ln}[ -\sqrt{-g}(\gamma^{\mu}\nabla_{\mu} + \sigma)]
  + {\cal O}(\frac{1}{N}) .
 \label{eqn:efact}
\end{equation}
{}From Eq.(\ref{eqn:efact}) we obtain effective potential in the leading
order of the $1/N$ expansion:
\begin{equation}
  V(\sigma) = \frac{1}{2\lambda_{0}}\sigma^{2}
  + \frac{ 1}{\int d^{D}\!x\,\sqrt{-g}}
  \mbox{Tr} \int_{0}^{\sigma} ds \,\sqrt{-g}S_{F}(x,y;s),
  \label{eqn:efpoten}
\end{equation}
where $\sigma$ is independent of the space-time coordinates and
 the potential $V(\sigma)$ is normalized so that $V(0)=0$
 with $S_{F}(x,y;s)$ defined by
\begin{equation}
  S_{F}(x,y;s) \equiv
  <x|\{-i\sqrt{-g}(\gamma^{\mu}\nabla_{\mu} + s)\}^{-1}|y> .
 \label{eqn:propdef}
\end{equation}
Through the definition (\ref{eqn:propdef}) we observe that the
two-point function $S_{F}(x,y;s)$ satisfies the following equation,
\begin{equation}
  (\gamma^{\mu}\nabla_{\mu} + s)S_{F}(x,y;s) =
  \frac{i}{\sqrt{-g}}\delta^{D}(x,y) ,
  \label{eqn:difeq}
\end{equation}
where $\delta^{D}(x,y)$ is Dirac's delta function
in curved space-time.
According to Eq.(\ref{eqn:difeq}) we may identify $S_{F}(x,y;s)$
to the propagator of the free massive fermion of mass $s$ in curved
space-time.

It is important to note here that the problem of calculating the
 effective potential for composite operator $\sigma$ in the
four-fermion interaction model in the leading order of the $1/N$
expansion reduces to that of finding the propagator of the massive
free fermion in curved space-time.
Fortunately in Einstein universe the two-point function $S_{F}(x,y;s)$ has
already been calculated in Ref.\cite{IIM}.
The resulting expression reads
\newpage
\begin{displaymath}
  S_{F}(x,y;s) = \int_{-\infty}^{\infty}
  \frac{d\omega}{2\pi}\, e^{i\omega(y^{0}-x^{0})}
  \frac{iK^{\frac{D-3}{2}}}{(4\pi)^{\frac{D-1}{2}}}
  \frac{\Gamma(\frac{D-1}{2}+i\beta)
  \Gamma(\frac{D-1}{2}-i\beta)}
  {\Gamma(\frac{D+1}{2})}
\end{displaymath}
\begin{equation}
  \times \left[ (s+i\gamma_{0}\omega)
  U(\mbox{\boldmath$x$},\mbox{\boldmath$y$})
  \cos(\frac{\sqrt{K}}{2}\sigma )
  \,\,_{2}F_{1}(\frac{D-1}{2}+i\beta,\frac{D-1}{2}-i\beta;\frac{D+1}{2};
  \cos^{2}\frac{\sqrt{K}}{2}\sigma ) \right.
 \label{eqn:propag}
\end{equation}
\begin{displaymath}
  + \left. \gamma^{i}n_{i}U(\mbox{\boldmath$x$},\mbox{\boldmath$y$})
  \frac{D-2}{2} \sin(\frac{\sqrt{K}}{2}\sigma )
  \,\,_{2}F_{1}(\frac{D-1}{2}+i\beta,\frac{D-1}{2}-i\beta;\frac{D-1}{2};
  \cos^{2}\frac{\sqrt{K}}{2}\sigma ) \right] ,
\end{displaymath}
where $K$ and $\beta$ are defined by
\begin{eqnarray}
  \beta & \equiv & \sqrt{\frac{s^{2}-\omega^{2}}{K}}, \\
 \mbox{and}\ \ \ \ R & \equiv & (D-1)(D-2)K , \label{eqn:curedef}
\end{eqnarray}
respectively where $R$ is the scalar curvature.
According to Eq.(\ref{eqn:propag}) we find that the trace of the
 two-point function $S_{F}(x,y;s)$ is given by
\begin{displaymath}
  \mbox{Tr}\sqrt{-g}S_{F}(x,y;s) = \int d^{D}x\sqrt{-g}\mbox{tr}\,S_{F}(x,x;s)
\end{displaymath}
\begin{equation}
  = \int d^{D}x\sqrt{-g}
  \int_{-\infty}^{\infty}\frac{d\omega}{2\pi}\,
  \frac{isK^{\frac{D-3}{2}}}{(4\pi)^{\frac{D-1}{2}}}
  \left|\frac{\Gamma(\frac{D-1}{2}+i\beta)}{\Gamma(1+i\beta)}\right|^{2}
  \Gamma(-\frac{D-3}{2}) \mbox{tr}\mbox{\boldmath$1$}, \label{eqn:greenfunc}
\end{equation}
where $\mbox{tr}\mbox{\boldmath$1$}$ is the trace of the unit Dirac matrix.
Inserting Eq.(\ref{eqn:greenfunc}) into
Eq.(\ref{eqn:efpoten}) we obtain the effective potential
in Einstein universe:
\begin{equation}
  V(\sigma) = \frac{\sigma^{2}}{2\lambda_{0}} +
  \int_{0}^{\sigma}\!\!\!\!ds\,\int_{-\infty}^{\infty}
  \frac{d\omega}{2\pi}\,
  \frac{isK^{\frac{D-3}{2}}}{(4\pi)^{\frac{D-1}{2}}}
  \left|\frac{\Gamma(\frac{D-1}{2}+i\beta)}{\Gamma(1+i\beta)}\right|^{2}
  \Gamma(-\frac{D-3}{2}) \mbox{tr}\mbox{\boldmath$1$}.
  \label{eqn:efpoteinnowick}
\end{equation}
In the second term of the right-hand side of
Eq.(\ref{eqn:efpoteinnowick})
, the path of the $\omega$ integral may be deformed by the Wick
rotation so that one can keep the path away from the poles appearing
in the Gamma functions.
By changing variable $\omega$ to $\omega^{'}$ through
 $\omega = i\omega^{'}$  we rewrite Eq.(\ref{eqn:efpoteinnowick})
 such that
\begin{equation}
  V(\sigma) = \frac{\sigma^{2}}{2\lambda_{0}} -
  \int_{0}^{\sigma}\!\!\!\!ds\,\int_{-\infty}^{\infty}
  \frac{d\omega^{'}}{2\pi}\,
  \frac{sK^{\frac{D-3}{2}}}{(4\pi)^{\frac{D-1}{2}}}
  \left|\frac{\Gamma(\frac{D-1}{2}+i\beta^{'})}
  {\Gamma(1+i\beta^{'})}\right|^{2}
  \Gamma(-\frac{D-3}{2}) \mbox{tr}\mbox{\boldmath$1$},
  \label{eqn:efpotein}
\end{equation}
where $\beta^{'}$ is defined by
$\beta^{'}\equiv\sqrt{\frac{s^{2}+\omega^{'2}}{K}}$.
For simplicity we shall omit the primes in $\omega^{'}$ and $\beta^{'}$
in the following.

The effective potential $V(\sigma)$ is in general divergent for
$D=2,3,4$.
Since the nature of the divergence is the same as that in the flat
(Minkowski) space-time, the divergence can be removed by the
usual flat-space renormalization for $D<4$. (Note that our model is
renormalizable for $D=2$ in usual sense and for $D=3$ in the sense of
the $1/N$ expansion \cite{D3}. For $D\geq 4$ the model is
 nonrenormalizable.)
The divergence associated with the vacuum energy may be
curvature-dependent. This divergence, however, is absent because of
 our normalization of the effective potential, $V(0)=0$, in
Eq.(\ref{eqn:efpotein}).
Note that in the leading order of the $1/N$ expansion the divergence
at $D=3$ in the effective potential $V(\sigma)$ happens to be
 absent can be seen by carefully analyzing Eq.(\ref{eqn:efpoten})
in the neighborhood of $D=3$.

Expanding Eq.(\ref{eqn:efpotein}) asymptotically about $K=0$ (weak
curvature expansion) we find
\begin{equation}
V(\sigma) = V_{0}(\sigma) +
        \frac{\mbox{tr}\mbox{\boldmath$1$}}{(4\pi)^{\frac{D}{2}}}
        \Gamma(2-\frac{D}{2})\frac{1}{12}
        \frac{R}{D-2}\sigma^{D-2}
        + {\cal O}(R^{2}) ,
        \label{eqn:efpotwek}
\end{equation}
where $R$ is given in Eq.(\ref{eqn:curedef}) and $V_{0}(\sigma)$ is
the effective potential in flat space-time given by
\begin{equation}
  V_{0}(\sigma)=\frac{\sigma^{2}}{2\lambda_{0}}-
  \int_{0}^{\sigma}ds\int_{-\infty}^{\infty}\frac{d\omega}{2\pi}
  \frac{s}{(4\pi)^{\frac{D-1}{2}}}(\omega^{2}+s^{2})^{\frac{D-3}{2}}
  \Gamma(-\frac{D-3}{2}) \mbox{tr}\mbox{\boldmath$1$}.
  \label{eqn:efpotmin}
\end{equation}
Though the $\omega$ integration in Eq.(\ref{eqn:efpotmin}) can be
 performed analytically, we leave it for convenience in the following
argument. (Note that the expression of the weak curvature expansion
 (\ref{eqn:efpotwek}) for $D=3$ and $4$ is in agreement with the
results obtained in Refs.\cite{ESY} and \cite{TTS} respectively.)
We impose the renormalization condition
\begin{equation}
  \left. \frac{\partial^{2}V_{0}}{\partial\sigma^{2}}
  \right|_{\sigma=\sigma_{0}}
  = \frac{\sigma^{D-2}_{0}}{\lambda_{r}},
  \label{eqn:renocond}
\end{equation}
where $\lambda_{r}$ is the renormalized coupling
constant and $\sigma_{0}$ is the renormalization scale.
Adopting the condition (\ref{eqn:renocond}) to the effective potential
(\ref{eqn:efpotmin}) we find that the renormalized coupling constant
$\lambda_{r}$ satisfies the following relation,
\begin{equation}
  \frac{1}{\lambda_{0}} =
  \frac{\sigma_{0}^{D-2}}{\lambda_{r}}
  + \int_{-\infty}^{\infty}\frac{d\omega}{2\pi} \,
  \frac{1}{(4\pi)^{\frac{D-1}{2}}}
  \left\{ \omega^{2}+(D-2)\sigma_{0}^{2} \right\}
  (\omega^{2}+\sigma_{0}^{2})^{\frac{D-5}{2}}
  \Gamma(-\frac{D-3}{2}) \mbox{tr}\mbox{\boldmath$1$} .
  \label{eqn:renocoupl}
\end{equation}
Inserting Eq.(\ref{eqn:renocoupl}) into Eq.(\ref{eqn:efpotmin})
we obtain the renormalized effective potential in the Minkowski
space-time:
\begin{displaymath}
  V_{0}(\sigma) =
  \frac{1}{2}\left[ \frac{\sigma_{0}^{D-2}}{\lambda_{r}}
  + \int_{-\infty}^{\infty}\frac{d\omega}{2\pi} \,
  \frac{1}{(4\pi)^{\frac{D-1}{2}}}
  \left\{ \omega^{2}+(D-2)\sigma_{0}^{2} \right\}
  (\omega^{2}+\sigma_{0}^{2})^{\frac{D-5}{2}}
  \Gamma(-\frac{D-3}{2}) \mbox{tr}\mbox{\boldmath$1$} \right]\sigma^{2}
  \nonumber
\end{displaymath}
\begin{equation}
  - \int_{0}^{\sigma}ds\int_{-\infty}^{\infty}\frac{d\omega}{2\pi}
  \frac{s}{(4\pi)^{\frac{D-1}{2}}}(\omega^{2}+s^{2})^{\frac{D-3}{2}}
  \Gamma(-\frac{D-3}{2}) \mbox{tr}\mbox{\boldmath$1$}.
\end{equation}
The nontrivial solution of the gap equation
$ \left.\frac{\partial V_{0}}{\partial\sigma}\right|_{\sigma=m_{0}}=0 $
provides us with the dynamical fermion mass $m_{0}$
 in the Minkowski space-time.
 By observing the gap equation we recognize that
above the critical coupling $\lambda_{cr}$ given by
\begin{equation}
  \frac{1}{\lambda_{cr}}\equiv \frac{1-D}{(4\pi)^{\frac{D}{2}}}
  \Gamma(1-\frac{D}{2})\mbox{tr}\mbox{\boldmath$1$} \label{eqn:defrenocou}
\end{equation}
the dynamical fermion mass $m_{0}$ is generated where
\begin{equation}
    m_{0}  =  \left\{
\frac{(4\pi)^{\frac{D}{2}}}{\Gamma(1-\frac{D}{2})\mbox{tr}\mbox{\boldmath$1$}}
    \left( \frac{1}{\lambda_{r}}-\frac{1}{\lambda_{cr}}\right)
    \right\}^{\frac{1}{D-2}}\sigma_{0}
    ~~~~~ \mbox{for $\lambda_{r} > \lambda_{cr}$}.
 \label{eqn:dynfermassmin}
\end{equation}
Below the critical coupling $\lambda_{cr}$ the fermions remain
massless: $m_{0}=0$.

Inserting Eq.(\ref{eqn:renocoupl}) into Eq.(\ref{eqn:efpotein})
we obtain a renormalized expression of the effective potential in
 Einstein universe :
\begin{displaymath}
  V(\sigma) =
  \frac{1}{2}\left[\frac{\sigma_{0}^{D-2}}{\lambda_{r}}
  +\int_{-\infty}^{\infty}\frac{d\omega}{2\pi}
  \frac{1}{(4\pi)^{\frac{D-1}{2}}}
  \left\{\omega^{2}+(D-2)\sigma_{0}^{2}\right\}
  (\omega^{2}+\sigma_{0}^{2})^{\frac{D-5}{2}}\Gamma(-\frac{D-3}{2})
  \mbox{tr}\mbox{\boldmath$1$}\right]\sigma^{2}
\end{displaymath}
\begin{equation}
  - \int_{0}^{\sigma}\!\!\!ds
  \,\int_{-\infty}^{\infty}\frac{d\omega}{2\pi}
  \frac{sK^{\frac{D-3}{2}}}{(4\pi)^{\frac{D-1}{2}}}
  \left|
  \frac{\Gamma(\frac{D-1}{2}+i\beta)}{\Gamma(1+i\beta)}
  \right|^{2}\Gamma(-\frac{D-3}{2})\mbox{tr}\mbox{\boldmath$1$}.
  \label{eqn:effoptrenoein}
\end{equation}
\begin{figure}[H]
\setlength{\unitlength}{0.240900pt}
\begin{picture}(1500,900)(-90,0)
\tenrm
\thinlines \drawline[-50](220,240)(1436,240)
\thicklines \path(220,134)(240,134)
\thicklines \path(1436,134)(1416,134)
\put(198,134){\makebox(0,0)[r]{-0.05}}
\thicklines \path(220,240)(240,240)
\thicklines \path(1436,240)(1416,240)
\put(198,240){\makebox(0,0)[r]{0}}
\thicklines \path(220,346)(240,346)
\thicklines \path(1436,346)(1416,346)
\put(198,346){\makebox(0,0)[r]{0.05}}
\thicklines \path(220,453)(240,453)
\thicklines \path(1436,453)(1416,453)
\put(198,453){\makebox(0,0)[r]{0.1}}
\thicklines \path(220,559)(240,559)
\thicklines \path(1436,559)(1416,559)
\put(198,559){\makebox(0,0)[r]{0.15}}
\thicklines \path(220,665)(240,665)
\thicklines \path(1436,665)(1416,665)
\put(198,665){\makebox(0,0)[r]{0.2}}
\thicklines \path(220,771)(240,771)
\thicklines \path(1436,771)(1416,771)
\put(198,771){\makebox(0,0)[r]{0.25}}
\thicklines \path(220,877)(240,877)
\thicklines \path(1436,877)(1416,877)
\put(198,877){\makebox(0,0)[r]{0.3}}
\thicklines \path(220,113)(220,133)
\thicklines \path(220,877)(220,857)
\put(220,68){\makebox(0,0){0}}
\thicklines \path(355,113)(355,133)
\thicklines \path(355,877)(355,857)
\put(355,68){\makebox(0,0){0.2}}
\thicklines \path(490,113)(490,133)
\thicklines \path(490,877)(490,857)
\put(490,68){\makebox(0,0){0.4}}
\thicklines \path(625,113)(625,133)
\thicklines \path(625,877)(625,857)
\put(625,68){\makebox(0,0){0.6}}
\thicklines \path(760,113)(760,133)
\thicklines \path(760,877)(760,857)
\put(760,68){\makebox(0,0){0.8}}
\thicklines \path(896,113)(896,133)
\thicklines \path(896,877)(896,857)
\put(896,68){\makebox(0,0){1}}
\thicklines \path(1031,113)(1031,133)
\thicklines \path(1031,877)(1031,857)
\put(1031,68){\makebox(0,0){1.2}}
\thicklines \path(1166,113)(1166,133)
\thicklines \path(1166,877)(1166,857)
\put(1166,68){\makebox(0,0){1.4}}
\thicklines \path(1301,113)(1301,133)
\thicklines \path(1301,877)(1301,857)
\put(1301,68){\makebox(0,0){1.6}}
\thicklines \path(1436,113)(1436,133)
\thicklines \path(1436,877)(1436,857)
\put(1436,68){\makebox(0,0){1.8}}
\thicklines \path(220,113)(1436,113)(1436,877)(220,877)(220,113)
\put(155,945){\makebox(0,0)[l]{\shortstack{$\left.
 \frac{\sqrt{\pi}(4\pi)^{\frac{D}{2}}V}
 {2\Gamma(-\frac{D-3}{2}) tr\mbox{\boldmath$1$}\sigma_{0}^{D}}
 \right|_{D=2}$}}}
\put(828,23){\makebox(0,0){$\sigma/\sigma_{0}$}}
\put(1166,336){\makebox(0,0)[l]{$K=0.5K_{cr}$}}
\put(909,495){\makebox(0,0)[l]{$K=K_{cr}$}}
\put(693,718){\makebox(0,0)[l]{$K=2K_{cr}$}}
\thinlines
\path(220,240)(220,240)(222,240)(224,240)(225,240)(227,240)(229,240)
(231,240)(234,240)(238,240)(241,240)(248,240)(255,239)(262,239)
(276,238)(304,235)(333,230)(389,219)(445,204)(501,187)(558,170)
(614,153)(642,146)(670,139)(699,134)(713,132)(727,130)(734,130)
(741,129)(748,129)(755,128)(758,128)(762,128)(764,128)(765,128)
(767,128)(769,128)(771,128)(772,128)(774,128)(776,128)(778,128)
(779,128)(781,128)(783,128)(786,128)(790,128)(794,128)(797,128)
(804,129)(811,129)
\thinlines
\path(811,129)(825,131)(839,133)(867,140)(896,149)(924,161)(952,176)
(1008,216)(1064,270)(1121,340)(1177,428)(1233,533)(1290,658)
(1346,803)(1371,877)
\thinlines
\path(220,240)(220,240)(222,240)(224,240)(225,240)(227,240)(229,240)
(231,240)(232,240)(234,240)(236,240)(238,240)(239,240)(241,240)
(245,240)(246,240)(248,240)(252,240)(255,240)(259,240)(262,240)
(266,240)(269,240)(273,240)(276,240)(283,240)(290,240)(297,240)
(304,240)(311,240)(319,240)(326,240)(333,240)(347,241)(361,241)
(375,241)(389,241)(403,241)(417,242)(431,242)(445,242)(473,244)
(501,245)(530,248)(558,251)(586,255)(614,260)(642,266)(670,273)
(727,292)(783,317)
\thinlines
\path(783,317)(839,351)(896,393)(952,447)(1008,512)(1064,591)(1121,683)
(1177,792)(1215,877)
\thinlines
\path(220,240)(220,240)(222,240)(224,240)(225,240)(227,240)(229,240)
(231,240)(234,240)(238,241)(241,241)(248,241)(255,241)(262,242)
(276,243)(290,244)(304,246)(333,251)(361,256)(389,264)(445,282)
(501,306)(558,337)(614,375)(670,420)(727,473)(783,535)(839,606)
(896,688)(952,781)(1003,877)
\end{picture}
 \begin{center}
    (a) Behavior of the effective potential
    for $D=2$ and $\lambda_{r}=3\lambda_{cr}$
  \end{center}
\end{figure}
\begin{figure}[H]
\setlength{\unitlength}{0.240900pt}
\begin{picture}(1500,900)(-90,0)
\tenrm
\thinlines \drawline[-50](220,222)(1436,222)
\thicklines \path(220,877)(240,877)
\thicklines \path(1436,877)(1416,877)
\put(198,877){\makebox(0,0)[r]{-0.006}}
\thicklines \path(220,768)(240,768)
\thicklines \path(1436,768)(1416,768)
\put(198,768){\makebox(0,0)[r]{-0.005}}
\thicklines \path(220,659)(240,659)
\thicklines \path(1436,659)(1416,659)
\put(198,659){\makebox(0,0)[r]{-0.004}}
\thicklines \path(220,550)(240,550)
\thicklines \path(1436,550)(1416,550)
\put(198,550){\makebox(0,0)[r]{-0.003}}
\thicklines \path(220,440)(240,440)
\thicklines \path(1436,440)(1416,440)
\put(198,440){\makebox(0,0)[r]{-0.002}}
\thicklines \path(220,331)(240,331)
\thicklines \path(1436,331)(1416,331)
\put(198,331){\makebox(0,0)[r]{-0.001}}
\thicklines \path(220,222)(240,222)
\thicklines \path(1436,222)(1416,222)
\put(198,222){\makebox(0,0)[r]{0}}
\thicklines \path(220,113)(240,113)
\thicklines \path(1436,113)(1416,113)
\put(198,113){\makebox(0,0)[r]{0.001}}
\thicklines \path(220,113)(220,133)
\thicklines \path(220,877)(220,857)
\put(220,68){\makebox(0,0){0}}
\thicklines \path(372,113)(372,133)
\thicklines \path(372,877)(372,857)
\put(372,68){\makebox(0,0){0.2}}
\thicklines \path(524,113)(524,133)
\thicklines \path(524,877)(524,857)
\put(524,68){\makebox(0,0){0.4}}
\thicklines \path(676,113)(676,133)
\thicklines \path(676,877)(676,857)
\put(676,68){\makebox(0,0){0.6}}
\thicklines \path(828,113)(828,133)
\thicklines \path(828,877)(828,857)
\put(828,68){\makebox(0,0){0.8}}
\thicklines \path(980,113)(980,133)
\thicklines \path(980,877)(980,857)
\put(980,68){\makebox(0,0){1}}
\thicklines \path(1132,113)(1132,133)
\thicklines \path(1132,877)(1132,857)
\put(1132,68){\makebox(0,0){1.2}}
\thicklines \path(1284,113)(1284,133)
\thicklines \path(1284,877)(1284,857)
\put(1284,68){\makebox(0,0){1.4}}
\thicklines \path(1436,113)(1436,133)
\thicklines \path(1436,877)(1436,857)
\put(1436,68){\makebox(0,0){1.6}}
\thicklines \path(220,113)(1436,113)(1436,877)(220,877)(220,113)
\put(155,945){\makebox(0,0)[l]{\shortstack{$\left.
\frac{\sqrt{\pi}(4\pi)^{\frac{D}{2}}V}
{2\Gamma(-\frac{D-3}{2})tr\mbox{\boldmath$1$}\sigma_{0}^{D}}
\right|_{D=3}$}}}
\put(828,23){\makebox(0,0){$\sigma/\sigma_{0}$}}
\put(1170,277){\makebox(0,0)[l]{$K=0.5K_{cr}$}}
\put(904,473){\makebox(0,0)[l]{$K=K_{cr}$}}
\put(706,659){\makebox(0,0)[l]{$K=2K_{cr}$}}
\thinlines
\path(220,222)(220,222)(222,222)(224,222)(226,222)(228,222)(230,222)
(232,222)(236,222)(240,222)(244,222)(252,222)(260,221)(268,221)
(283,220)(315,218)(347,215)(410,207)(473,197)(537,186)(600,175)
(663,165)(695,161)(711,159)(727,157)(743,156)(758,155)(766,155)
(770,155)(774,155)(778,155)(782,155)(784,155)(786,154)(788,154)
(790,154)(792,154)(794,154)(796,154)(798,155)(802,155)(804,155)
(806,155)(814,155)(818,155)(822,155)(838,156)(845,157)(853,158)
(885,162)(901,166)
\thinlines
\path(901,166)(917,169)(948,179)(980,191)(1043,224)(1107,271)(1170,333)
(1233,414)(1297,515)(1360,637)(1423,783)(1436,817)
\thinlines\path(220,222)(220,222)(222,222)(224,222)(226,222)(228,222)
(230,222)(232,222)(236,222)(240,222)(244,222)(248,222)(252,222)
(260,222)(268,222)(275,222)(283,222)(291,222)(299,222)(307,222)
(315,222)(331,222)(339,222)(347,222)(363,222)(378,223)(394,223)
(410,223)(442,224)(458,224)(473,225)(505,226)(537,228)(568,230)
(600,234)(632,238)(663,243)(695,249)(727,257)(790,277)(853,305)
(917,342)(980,389)(1043,449)(1107,523)(1170,613)(1233,721)
(1297,850)(1308,877)
\thinlines
\path(220,222)(220,222)(222,222)(224,222)(226,222)(228,222)(230,222)
(232,222)(236,222)(240,222)(244,222)(252,223)(260,223)(268,224)
(283,225)(299,226)(315,228)(347,232)(378,238)(410,245)(473,264)
(537,288)(600,320)(663,359)(727,406)(790,463)(853,531)(917,610)
(980,703)(1043,810)(1078,877)
\end{picture}
  \begin{center}
    (b) Behavior of the effective potential
    for $D=3$ and $\lambda_{r}=3\lambda_{cr}$
  \end{center}
 \caption[figefpot]{Behavior of the effective potential
   as a function of $\sigma$ for some typical value of
   curvature $K$ in the case that
   $\lambda_{r} > \lambda_{cr}$.}
 \label{tbl:figefpot}
\end{figure}
\begin{figure}[H]
\setlength{\unitlength}{0.240900pt}
\begin{picture}(1500,900)(-90,0)
\tenrm
\thinlines \drawline[-50](220,222)(1436,222)
\thicklines \path(220,877)(240,877)
\thicklines \path(1436,877)(1416,877)
\put(198,877){\makebox(0,0)[r]{-0.006}}
\thicklines \path(220,768)(240,768)
\thicklines \path(1436,768)(1416,768)
\put(198,768){\makebox(0,0)[r]{-0.005}}
\thicklines \path(220,659)(240,659)
\thicklines \path(1436,659)(1416,659)
\put(198,659){\makebox(0,0)[r]{-0.004}}
\thicklines \path(220,550)(240,550)
\thicklines \path(1436,550)(1416,550)
\put(198,550){\makebox(0,0)[r]{-0.003}}
\thicklines \path(220,440)(240,440)
\thicklines \path(1436,440)(1416,440)
\put(198,440){\makebox(0,0)[r]{-0.002}}
\thicklines \path(220,331)(240,331)
\thicklines \path(1436,331)(1416,331)
\put(198,331){\makebox(0,0)[r]{-0.001}}
\thicklines \path(220,222)(240,222)
\thicklines \path(1436,222)(1416,222)
\put(198,222){\makebox(0,0)[r]{0}}
\thicklines \path(220,113)(240,113)
\thicklines \path(1436,113)(1416,113)
\put(198,113){\makebox(0,0)[r]{0.001}}
\thicklines \path(220,113)(220,133)
\thicklines \path(220,877)(220,857)
\put(220,68){\makebox(0,0){0}}
\thicklines \path(463,113)(463,133)
\thicklines \path(463,877)(463,857)
\put(463,68){\makebox(0,0){0.1}}
\thicklines \path(706,113)(706,133)
\thicklines \path(706,877)(706,857)
\put(706,68){\makebox(0,0){0.2}}
\thicklines \path(950,113)(950,133)
\thicklines \path(950,877)(950,857)
\put(950,68){\makebox(0,0){0.3}}
\thicklines \path(1193,113)(1193,133)
\thicklines \path(1193,877)(1193,857)
\put(1193,68){\makebox(0,0){0.4}}
\thicklines \path(1436,113)(1436,133)
\thicklines \path(1436,877)(1436,857)
\put(1436,68){\makebox(0,0){0.5}}
\thicklines \path(220,113)(1436,113)(1436,877)(220,877)(220,113)
\put(155,945){\makebox(0,0)[l]{\shortstack{$\left.
\frac{\sqrt{\pi}(4\pi)^{\frac{D}{2}}V}
{2\Gamma(-\frac{D-3}{2})tr\mbox{\boldmath$1$}\sigma_{0}^{D}}
\right|_{D=3}$}}}
\put(828,23){\makebox(0,0){$\sigma/\sigma_{0}$}}
\put(1023,331){\makebox(0,0)[l]{$K=0.5K_{cr}$}}
\put(1096,440){\makebox(0,0)[l]{$K=K_{cr}$}}
\put(901,768){\makebox(0,0)[l]{$K=2K_{cr}$}}
\thinlines
\path(220,222)(220,222)(226,222)(233,222)(239,222)(245,222)(252,223)
(258,223)(271,223)(283,224)(296,224)(321,226)(347,228)(372,231)
(423,238)(473,247)(524,258)(625,287)(727,323)(828,368)(1031,483)
(1233,633)(1436,818)(1436,818)
\thinlines
\path(220,222)(220,222)(226,222)(233,222)(239,222)(245,222)(252,223)
(258,223)(271,223)(283,224)(296,225)(321,227)(347,229)(372,232)
(423,240)(473,250)(524,262)(625,294)(727,334)(828,384)(1031,510)
(1233,674)(1436,877)(1436,877)
\thinlines
\path(220,222)(220,222)(226,222)(233,222)(239,222)(245,222)(252,223)
(258,223)(271,223)(283,224)(296,225)(321,227)(347,230)(372,234)
(423,243)(473,254)(524,268)(625,304)(727,350)(828,406)(1031,549)
(1233,735)(1360,877)
\thinlines \path(1120,353)(1120,353)(952,440)
\thinlines \path(1193,462)(1193,462)(1088,550)
\thinlines \path(1003,757)(1003,757)(1147,659)
\end{picture}
 \caption[figefpot2]{Behavior of the effective potential
   as a function of $\sigma$ for some typical value of curvature $K$
   in the case that $\lambda_{r} < \lambda_{cr}$ for $D=3$.}
 \label{tbl:figefpot2}
\end{figure}
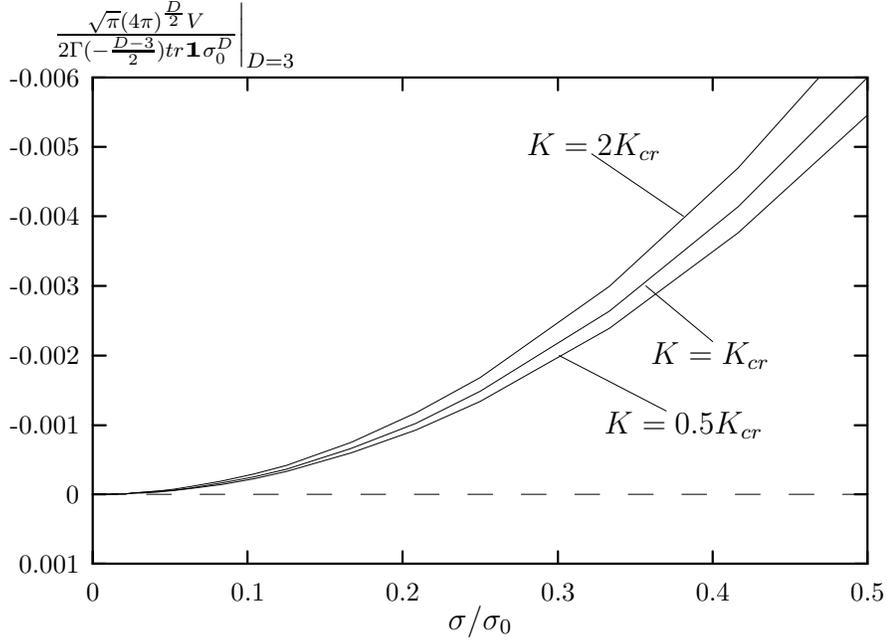
By the use of the expression of the effective potential as given above
we may study the behavior of the effective potential as a function of
$\sigma$ through the numerical integrations.
 In the numerical integration in $\omega$ we introduced a suitable
upper and lower bound and performed the numerical integration between
these bounds.
 By assuming the sufficiently large absolute value of these bounds and
checking the stability of the integral under the change of the bounds
we obtained the numerical value for $V(\sigma)$ for each value of
$\sigma$ with $\lambda_{r}$ and $K$ kept fixed.

 In Fig.\ref{tbl:figefpot} the behavior of the effective potential
$V(\sigma)$ as a function of $\sigma$ is plotted for the case
$\lambda_{r} > \lambda_{cr}$ (as a typical example we choose
$\lambda_{r}=3 \lambda_{cr}$) for which the chiral symmetry is broken
and the fermion mass is generated in the flat space  where $K=0$. The
behavior of the effective potential is shown only for $D=2$ and $3$.
We clearly observe that the behavior of the effective potential
$V(\sigma)$ for $K=0$ which is typical of the broken phase is
gradually changed into the typical behavior in the symmetric phase as
we increase the curvature $K$. The critical curvature at which the
symmetry restoration takes place can be calculated analytically as will
be shown later. In Fig.\ref{tbl:figefpot2} the behavior of $V(\sigma)$
is presented for the case $\lambda_{r} \leq \lambda_{cr}$ (as a typical
example we choose $\lambda_{r} = \lambda_{cr} / 2$ and the
space-time dimension $D=3$). We observe that the symmetric phase stays
symmetric if we change the curvature.

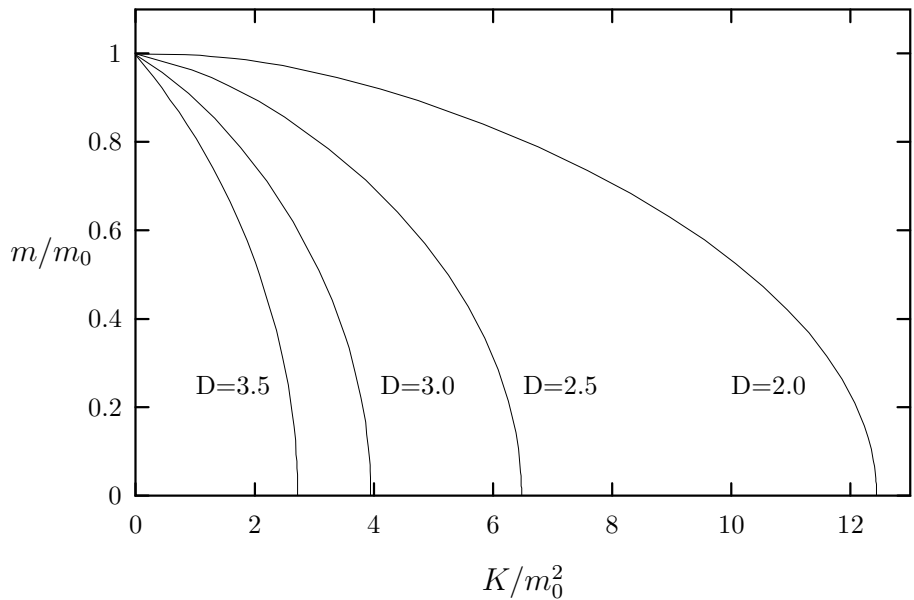
\begin{figure}[H]
  \begin{center}
\setlength{\unitlength}{0.240900pt}
\begin{picture}(1500,900)(90,0)
\tenrm
\thicklines \path(220,113)(240,113)
\thicklines \path(1436,113)(1416,113)
\put(198,113){\makebox(0,0)[r]{0}}
\thicklines \path(220,252)(240,252)
\thicklines \path(1436,252)(1416,252)
\put(198,252){\makebox(0,0)[r]{0.2}}
\thicklines \path(220,391)(240,391)
\thicklines \path(1436,391)(1416,391)
\put(198,391){\makebox(0,0)[r]{0.4}}
\thicklines \path(220,530)(240,530)
\thicklines \path(1436,530)(1416,530)
\put(198,530){\makebox(0,0)[r]{0.6}}
\thicklines \path(220,669)(240,669)
\thicklines \path(1436,669)(1416,669)
\put(198,669){\makebox(0,0)[r]{0.8}}
\thicklines \path(220,808)(240,808)
\thicklines \path(1436,808)(1416,808)
\put(198,808){\makebox(0,0)[r]{1}}
\thicklines \path(220,113)(220,133)
\thicklines \path(220,877)(220,857)
\put(220,68){\makebox(0,0){0}}
\thicklines \path(407,113)(407,133)
\thicklines \path(407,877)(407,857)
\put(407,68){\makebox(0,0){2}}
\thicklines \path(594,113)(594,133)
\thicklines \path(594,877)(594,857)
\put(594,68){\makebox(0,0){4}}
\thicklines \path(781,113)(781,133)
\thicklines \path(781,877)(781,857)
\put(781,68){\makebox(0,0){6}}
\thicklines \path(968,113)(968,133)
\thicklines \path(968,877)(968,857)
\put(968,68){\makebox(0,0){8}}
\thicklines \path(1155,113)(1155,133)
\thicklines \path(1155,877)(1155,857)
\put(1155,68){\makebox(0,0){10}}
\thicklines \path(1342,113)(1342,133)
\thicklines \path(1342,877)(1342,857)
\put(1342,68){\makebox(0,0){12}}
\thicklines \path(220,113)(1436,113)(1436,877)(220,877)(220,113)
\put(23,495){\makebox(0,0)[l]{\shortstack{$m/m_{0}$}}}
\put(828,-22){\makebox(0,0){$K/m_{0}^{2}$}}
\put(1155,287){\makebox(0,0)[l]{D=2.0}}
\put(828,287){\makebox(0,0)[l]{D=2.5}}
\put(604,287){\makebox(0,0)[l]{D=3.0}}
\put(314,287){\makebox(0,0)[l]{D=3.5}}
\thinlines
\path(1383,113)(1383,113)(1383,114)(1383,115)(1383,116)(1383,118)
(1383,120)(1383,122)(1383,124)(1383,127)(1383,131)(1382,136)
(1382,140)(1381,150)(1380,159)(1378,168)(1375,186)(1370,204)
(1364,223)(1349,259)(1329,296)(1305,332)(1277,369)(1243,405)
(1205,442)(1162,478)(1113,515)(1058,551)(997,588)(929,624)
(852,661)(765,697)(663,734)(604,752)(536,770)(496,779)(451,789)
(424,793)(393,798)(374,800)(351,802)(338,803)(321,805)(297,806)
(221,807)
\thinlines
\path(826,113)(826,113)(826,115)(826,116)(826,118)(826,119)(826,121)
(826,122)(826,125)(825,128)(825,132)(825,138)(825,144)(824,150)
(823,163)(822,175)(821,187)(817,212)(811,237)(805,262)(789,311)
(768,361)(742,410)(711,460)(674,509)(631,559)(581,609)(523,658)
(454,708)(413,733)(366,757)(339,770)(308,782)(221,807)
\thinlines
\path(221,805)(221,805)(262,778)(303,745)(344,706)(385,660)(426,607)
(467,544)(508,467)(529,420)(549,363)(555,345)(565,306)(574,267)
(581,229)(583,209)(585,190)(587,171)(588,161)(588,152)(589,142)
(589,137)(589,132)(589,127)(589,125)(589,123)(589,120)(589,118)
(589,117)(589,115)(589,114)(589,113)
\thinlines
\path(221,804)(221,804)(234,788)(248,771)(261,754)(274,735)(288,716)
(301,696)(315,674)(328,652)(341,628)(355,602)(368,575)(381,546)
(395,514)(408,479)(415,460)(441,373)(451,330)(460,287)(466,243)
(469,222)(471,200)(472,178)(473,167)(473,156)(474,146)(474,140)
(474,135)(474,129)(474,124)(474,121)(474,118)(474,116)(474,113)
\end{picture}
    \caption[figgapeq]{Behavior of the dynamical fermion mass
    as a function of curvature $K$ for $D=2$,~$2.5$,~$3$ and $3.5$.}
    \label{tbl:figgapeq}
  \end{center}
\end{figure}

The dynamical fermion mass $m$ in Einstein universe is calculated by
solving the gap equation
$ \left.\frac{\partial V}{\partial\sigma}\right|_{\sigma=m}=0 $:
\begin{eqnarray}
  \frac{\sigma_{0}^{D-2}}{\lambda_{r}}
  & + & \int_{-\infty}^{\infty}\frac{d\omega}{2\pi}
  \frac{1}{(4\pi)^{\frac{D-1}{2}}} \Biggl[
  \left\{ \omega^{2}+(D-2)\sigma_{0}^{2}\right\}
  (\omega^{2}+\sigma_{0}^{2})^{\frac{D-5}{2}}  \nonumber \\
  & - &
  \left. K^{\frac{D-3}{2}}\left|\frac{\Gamma(\frac{D-1}{2}+i\beta)}
  {\Gamma(1+i\beta)}\right|^{2} \,\, \right]
  \Gamma(-\frac{D-3}{2})\mbox{tr}\mbox{\boldmath$1$} =0 ,
  \label{eqn:gapeqein}
\end{eqnarray}
where $\beta = \sqrt{\frac{m^{2}+\omega^{2}}{K}}$.
Fig.\ref{tbl:figgapeq} represents the behavior of the dynamical
fermion mass which is obtained by solving Eq.(\ref{eqn:gapeqein})
 numerically for $D=2.0,2.5,3.0$ and $3.5$.
The numerical integration in Eq.(\ref{eqn:gapeqein}) was
performed in the same way as in Fig.\ref{tbl:figefpot} and
Fig.\ref{tbl:figefpot2}.
As we have observed in Fig.\ref{tbl:figefpot} and
Fig.\ref{tbl:figgapeq}, the phase transition caused by the change of
the curvature is of the second order. This means that the critical
curvature may be calculated by taking the limit $m \rightarrow 0$ in
the gap equation (\ref{eqn:gapeqein}). We let $m \rightarrow 0$
in Eq.(\ref{eqn:gapeqein}) and perform the integration analytically.
 We find
\begin{equation}
  \frac{1}{\lambda_{r}}+\frac{D-1}{(4\pi)^{\frac{D}{2}}}
  \Gamma(1-\frac{D}{2})\mbox{tr}\mbox{\boldmath$1$}-
  \left(\frac{K_{cr}}{\sigma_{0}^{2}}\right)^{\frac{D-2}{2}}
  \frac{1}{\sqrt{\pi}(4\pi)^{\frac{D}{2}}}
  \Gamma(\frac{D-1}{2})\Gamma(\frac{D}{2})\Gamma(1-\frac{D}{2})
  \mbox{tr}\mbox{\boldmath$1$}=0.
  \label{eqn:gapeqm0}
\end{equation}
Taking into account Eq.(\ref{eqn:dynfermassmin}) with
Eq.(\ref{eqn:defrenocou}) we rewrite Eq.(\ref{eqn:gapeqm0}) in the
following form,
\begin{equation}
  \left[ m_{0}^{D-2}-K_{cr}^{\frac{D-2}{2}}\frac{1}{\sqrt{\pi}}
  \Gamma(\frac{D-1}{2})\Gamma(\frac{D}{2})\right]
  \Gamma(1-\frac{D}{2})=0.
  \label{eqn:critgap}
\end{equation}
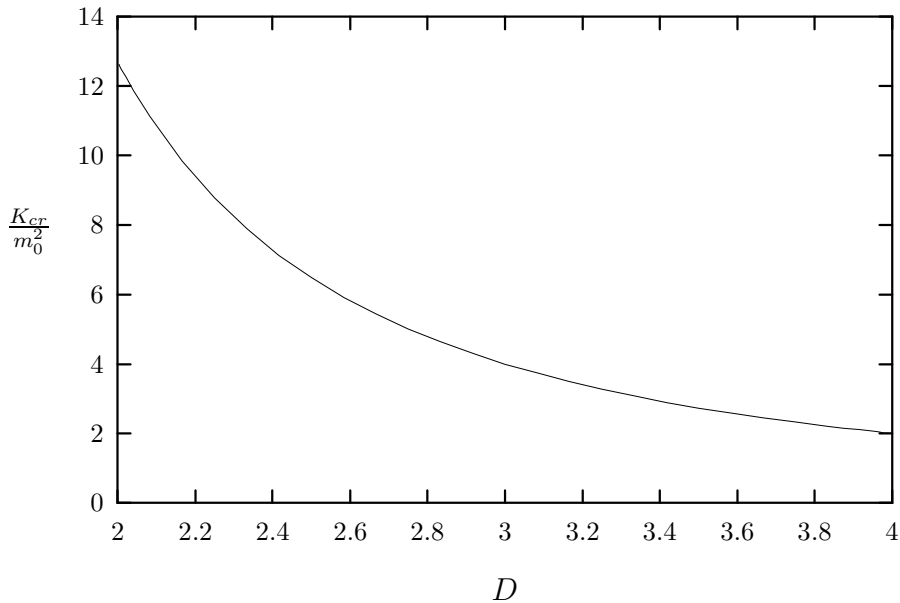
\begin{figure}[H]
  \begin{center}
\setlength{\unitlength}{0.240900pt}
\begin{picture}(1500,900)(90,0)
\tenrm
\thicklines \path(220,113)(240,113)
\thicklines \path(1436,113)(1416,113)
\put(198,113){\makebox(0,0)[r]{0}}
\thicklines \path(220,222)(240,222)
\thicklines \path(1436,222)(1416,222)
\put(198,222){\makebox(0,0)[r]{2}}
\thicklines \path(220,331)(240,331)
\thicklines \path(1436,331)(1416,331)
\put(198,331){\makebox(0,0)[r]{4}}
\thicklines \path(220,440)(240,440)
\thicklines \path(1436,440)(1416,440)
\put(198,440){\makebox(0,0)[r]{6}}
\thicklines \path(220,550)(240,550)
\thicklines \path(1436,550)(1416,550)
\put(198,550){\makebox(0,0)[r]{8}}
\thicklines \path(220,659)(240,659)
\thicklines \path(1436,659)(1416,659)
\put(198,659){\makebox(0,0)[r]{10}}
\thicklines \path(220,768)(240,768)
\thicklines \path(1436,768)(1416,768)
\put(198,768){\makebox(0,0)[r]{12}}
\thicklines \path(220,877)(240,877)
\thicklines \path(1436,877)(1416,877)
\put(198,877){\makebox(0,0)[r]{14}}
\thicklines \path(220,113)(220,133)
\thicklines \path(220,877)(220,857)
\put(220,68){\makebox(0,0){2}}
\thicklines \path(342,113)(342,133)
\thicklines \path(342,877)(342,857)
\put(342,68){\makebox(0,0){2.2}}
\thicklines \path(463,113)(463,133)
\thicklines \path(463,877)(463,857)
\put(463,68){\makebox(0,0){2.4}}
\thicklines \path(585,113)(585,133)
\thicklines \path(585,877)(585,857)
\put(585,68){\makebox(0,0){2.6}}
\thicklines \path(706,113)(706,133)
\thicklines \path(706,877)(706,857)
\put(706,68){\makebox(0,0){2.8}}
\thicklines \path(828,113)(828,133)
\thicklines \path(828,877)(828,857)
\put(828,68){\makebox(0,0){3}}
\thicklines \path(950,113)(950,133)
\thicklines \path(950,877)(950,857)
\put(950,68){\makebox(0,0){3.2}}
\thicklines \path(1071,113)(1071,133)
\thicklines \path(1071,877)(1071,857)
\put(1071,68){\makebox(0,0){3.4}}
\thicklines \path(1193,113)(1193,133)
\thicklines \path(1193,877)(1193,857)
\put(1193,68){\makebox(0,0){3.6}}
\thicklines \path(1314,113)(1314,133)
\thicklines \path(1314,877)(1314,857)
\put(1314,68){\makebox(0,0){3.8}}
\thicklines \path(1436,113)(1436,133)
\thicklines \path(1436,877)(1436,857)
\put(1436,68){\makebox(0,0){4}}
\thicklines \path(220,113)(1436,113)(1436,877)(220,877)(220,113)
\put(45,540){\makebox(0,0)[l]{\shortstack{$\frac{K_{cr}}{m_{0}^{2}}$}}}
\put(828,-22){\makebox(0,0){$D$}}
\thinlines
\path(222,802)(222,802)(223,800)(226,794)(233,782)(245,760)(271,720)
(321,650)(372,592)(423,544)(473,502)(524,467)(575,436)(625,410)
(676,386)(727,366)(777,348)(828,331)(879,317)(929,304)(980,292)
(1031,281)(1081,271)(1132,262)(1183,254)(1233,247)(1284,240)
(1335,233)(1360,230)(1385,228)(1411,225)(1417,224)(1423,223)
(1427,223)(1430,223)(1431,223)(1433,222)(1434,222)
\end{picture}
    \caption[figdvsk]{Behavior of the critical curvature
    $K_{cr}$ as a function of dimension $D$.}
    \label{tbl:figdvskcr}
  \end{center}
\end{figure}

In Fig.\ref{tbl:figdvskcr} we show dependence of the critical
curvature $K_{cr}$ to dimension $D$.
We see from Eq.(\ref{eqn:critgap}) that for some special values of
$D$ $K_{cr}$ is given by
\begin{equation}
  \left\{
  \begin{array}{cccc}
    K_{cr} & = & 4e^{2\gamma_{E}}m_{0}^{2} & \mbox{at $D=2.0$}, \\
    K_{cr} & = &
    \left(\frac{8}{\pi}\right)^{2}m_{0}^{2} & \mbox{at $D=2.5$}, \\
    K_{cr} & = & 4m_{0}^{2} & \mbox{at $D=3.0$},\\
    K_{cr} & = &
    \left(\frac{16}{3\sqrt{2\pi}}\right)^{\frac{4}{3}}m_{0}^{2}
     & \mbox{at $D=3.5$}, \\
    K_{cr} & = & 2m_{0}^{2} & \mbox{at $D=4.0$},
  \end{array}
  \right.
  \label{eqn:tblofkcr}
\end{equation}
where $\gamma_{E}$ is the Euler constant.

In Eq.(\ref{eqn:tblofkcr}) the critical curvature $K_{cr}$ for $D=2$
reproduces the result obtained in Ref.\cite{SWKJ}.
It is well-known that in the case of the two dimensional space-time
the field theory in $R\otimes S$ that is regarded as an
Euclidean analog of the Einstein space-time is equivalent to the
 finite-temperature field theory.
Since the effective potential for the two-dimensional
finite-temperature four-fermion theory is known \cite{Temp},
 we compare our result
(\ref{eqn:efpotein}) with the previous finite-temperature result.
In the two-dimensional limit $D\rightarrow 2$ the effective potential
(\ref{eqn:efpotein}) reduces to
\begin{eqnarray}
 V(\sigma)&\rightarrow&\frac{\sigma^{2}}{2\lambda_{0}}
        -\int_{0}^{\sigma}ds\int_{-\infty}^{\infty}\frac{d\omega}{3\pi}
         \frac{s}{\sqrt{K}}\frac{1}{\beta}\tanh \pi\beta \\
      &=&\frac{\sigma^{2}}{2\lambda_{0}}-\frac{\sqrt{K}}{2\pi}
   \sum_{n=-\infty}^{\infty}(\sqrt{\omega_{n}^{2}+\sigma^{2}}
   -\sqrt{\omega_{n}^{2}}),
  \label{eqn:efpotfintemp}
\end{eqnarray}
where $\omega_{n}=\frac{2n+1}{2}\sqrt{K}$. Here we utilized the
 formula,
\begin{equation}
 \sum_{n=1}^{\infty}\frac{1}{x^{2}+(2n-1)^{2}}
   = \frac{\pi}{4x}\tanh\pi\frac{x}{2}.
\end{equation}
With recourse to the following relation between the curvature
$K$ and the temperature $T$
\begin{equation}
  \frac{1}{k_{B}T}=\frac{2\pi}{\sqrt{K}},
\end{equation}
with $k_{B}$ the Boltzmann constant, we realize that
 the effective potential (\ref{eqn:efpotein}) in the Einstein universe
 for $D=2$ is in agreement with that of the finite-temperature
four-fermion theory in $D=2$ and the result for $D=2$ given in
Eq.(\ref{eqn:tblofkcr}) is consistent with the well-known formula of
the critical temperature for the four-fermion interaction model at
finite-temperature (See Ref.\cite{Temp}).

We have found the presence of the curvature-induced dynamical symmetry
 restoration for the system of self-interacting fermions in Einstein
universe in the leading order of the $1/N$ expansion. The
dynamical fermion mass is found to
disappear through the second-order phase transition. The critical
curvature at which the symmetry restoration takes place is calculated
analytically.

Since Einstein universe is the static universe, our
result in the present paper can not be applied directly to the
evolution of the universe in its early stage. Our model, however,
gives a hint on the scenario of the symmetry breaking in the early
universe. In order to make our result more realistic it is necessary
for us to extend our analysis to the more general universe such as the
Robertson-Walker universe and study the Einstein equation with the
energy-momentum tensor calculated in the underlying theory. It is our
hope to give a fundamental contribution to the understanding of the
dynamical origin of the inflationary expansion of the universe.

\subsection*{Acknowledgments}

The authors would like to thank K.~Fukazawa, S.~Mukaigawa and H.~Sato
for useful conversations. One of the authors (T.~M.) is supported
financially in part by the Monbusho Grant-in-Aid for Scientific
Research (A) under Contract No.07404002.

\end{document}